\documentclass[preprint,pra,aps,epsfig,showpacs]{revtex4}

\usepackage{graphicx}

\newcommand{\beq}{\begin{eqnarray}}
\newcommand{\eeq}{\end{eqnarray}}

\begin{document}
\setcounter{page}{1}

\title{State diagram for continuous quasi-one dimensional systems in
optical lattices}
\author{C. Carbonell-Coronado,  F.~De~Soto  and M.C. Gordillo}
\affiliation{Departamento de Sistemas F\'{\i}sicos Qu\'{\i}micos y Naturales,
Universidad Pablo de Olavide. 41013 Sevilla, Spain}

\begin{abstract}
We studied the appearance of Mott insulator domains of  
hard sphere bosons on quasi one-dimensional optical lattices when  
an harmonic trap was superimposed along the main axis of the system. 
Instead of the standard approximation represented by the Bose-Hubbard model, 
we described those arrangements by continuous Hamiltonians that depended 
on the same parameters as the experimental setups.   
We found that for a given trap the optical potential depth, $V_0$, needed to create a single connected Mott 
domain decreased with the number of atoms loaded on the lattice. If the confinement was large enough, it reached  
a minimum when, in absence of any optical lattice,  the atom density at the center of the trap was the equivalent of one particle
per optical well. For larger densities, the creation of that single domain proceeded
via an intermediate shell structure in which Mott domains alternated with superfluid ones.          

\end{abstract}


\maketitle

\section{Introduction}
The interference between one or several pairs of laser beams can be manipulated to produce a regularly 
varying light intensity pattern in a region of space. 
That creates an effective potential that can be felt by neutral atoms and whose most general form 
in three dimensions is \cite{gre2,bloch1,bloch2,greiner}  
\begin{equation} \label{pot}
V_{ext}(x_i,y_i,z_i) = V_x \sin^2(k_x x_i) +  
 V_y \sin^2(k_y y_i) + V_z \sin^2(k_z z_i) 
,
\end{equation}
expression that depends on the laser wavelengths $\lambda_x,\lambda_y,\lambda_z$ through   
$k_{x,y,z}=2\pi/\lambda_{x,y,z}$.     
The positions of the minima in Eq. (\ref{pot}) can be arranged to build different 
types of periodic (optical) lattices. The degree of confinement of the atoms in the lattice nodes can
be controlled by varying the intensity of the laser light and in that way the depths ($V_x,V_y$,$V_z$) of those wells.
Those three parameters can be changed independently, to produce asymmetric arrangements in which the atoms are 
more or less confined in particular directions.
In this context, a fairly common experimental setup makes  $V_x = V_y >> V_z$, and creates quasi-one dimensional tubes
in which the atoms move mainly along the principal axis of the cylinder \cite{paredes,stoferle,clement,haller,imai}. 
Since in most experiments  
additional trapping in the form of an
harmonic potential along the $z$ axis of the tubes is imposed, we can describe those systems by the following Hamiltonian: 
\begin{equation}\label{hamiltonian}
H = \sum_{i=1}^N \left[ -\frac{\hbar^2}{2m} \triangle  + V_{ext} (x_i,y_i,z_i) + \frac{1}{2} m \omega_z^2 z_i^2\right]  + \sum_{i<j} V(r_{ij}) 
\end{equation}
where $ V(r_{ij})$ represents the interatomic potential between the pair of atoms $i$ and $j$, located at a distance $r_{ij}$ from each other, 
$m$ is the mass of the atoms loaded in the optical lattice, and $\omega_z=2\pi f_z$, with $f_z$ the harmonic trapping longitudinal frequency.
When $V_x = V_y >> V_z$ the atoms are confined to an almost one-dimensional
cylinder and 
$V_{ext}(x_i,y_i,z_i)$ takes the approximate form:
\begin{equation} \label{pot1d}
V_{ext}(x,y,z) =   \frac{1}{2} m \omega_{\perp}^2 (x^2+y^2)  + V_0 \sin^2(k_z z).
\end{equation}
In this work, we solved the   
Sch\"odinger equation corresponding to the above Hamiltonian (Eq.(\ref{hamiltonian})). No simplification was involved beyond 
considering the interatomic potential to be of the hard spheres (HS) type. This means, 
$V(r_{ij})=+\infty$ for $r_{ij}<a$ and 
$V(r_{ij})= 0$ for $r_{ij}>a$, $a$ being the scattering length of the atoms. This interaction has been widely used   
both to describe homogeneous diluted gases 
\cite{boro1,gregori,blume,pro,boro2,rossi} and, to a lesser extend,
bosons loaded in optical lattices \cite{feli1,pilati,feli2,feli3,feli4}. 
   
Contrarily to the continuous treatment of the interactions that we propose, 
the standard approach to describe neutral atoms in optical lattices is the 
afforded by the discrete Bose-Hubbard (BH) Hamiltonian \cite{jaksch,bloch2},
a
discrete model obtained by simplifying  
Eq. (\ref{hamiltonian}).  
\begin{equation}
H=-J \sum_{<ij>} b_i^+ b_j+ \frac{U}{2} \sum_i n_i(n_i-1) +  \sum_i \epsilon_i n_i.
\label{hub}
\end{equation}
In this expression, the $i$'s label the positions of the minima of the optical lattice potential,
the only possible 
locations of the neutral atoms. 
Only the interactions with the nearest neighbor sites $j$ are considered (pairs $<ij>$ in Eq. (\ref{hub})).  
$b^+_i(b_i)$ is the creation (annihilation) operator for a boson at 
site $i$, and $n_i$ stands for the number of neutral atoms at that site.  
$J$, $U$ and $\epsilon_i$ are parameters related to the experimental ones ($V_0$, $\lambda$, $E_R$, $\omega_z$ and $\omega_{\perp}$) in a rather 
complicate way \cite{bloch2}. For a quasi-one dimensional system:
\begin{equation}\label{J}
 J=\frac{4}{\sqrt\pi}E_{R}\left(\frac{V_0}{E_R}\right)^{3/4}e^{-2\sqrt{\frac{V_0}{E_R}}},
\end{equation}
\begin{equation}\label{U}
 U=\sqrt{\frac{2}{\pi}}\hbar\omega_{\perp}\left(\frac{V_0}{E_R} \right)^{1/4}\frac{2\pi}{(\lambda/a)}.
\end{equation}
Here, $J$ is the hopping matrix element between nearest-neighbor sites, and $U$ represents the on site repulsion of two atoms located at the same potential 
minimum. Eq. (\ref{U}) was derived supposing that the interaction between those atoms was adequately described by a pseudopotential.   
$\epsilon_i$ is the energy offset at each potential well, and in our case takes into account the influence of
the harmonic trap along the main axis of the tube. This means \cite{jaksch,nandini}:
\begin{equation} \label{epsilon}
\epsilon_i \sim 1/2 m \omega_z^2 z_i^2 = V_c r_i^2 
\end{equation}
where $r_i=z_i$ ($r$ being the standard notation in BH Hamiltonians) is the longitudinal distance of the site $i$ to the center of 
the trap \cite{dm,dm2,dm3,dm,dm4}.     
One has also to bear in mind that 
Eqs. (\ref{J}) and (\ref{U}) are only valid when the potential wells are deep enough for the ground state to be described  
by a set of Wannier functions localized within that potential well, and when 
the energy difference between the ground and the first excited (Bloch) state of the complete Hamiltonian 
is much larger than the interparticle interaction of two atoms loaded on the same site. If all conditions above are fulfilled, 
and when $\epsilon_i$ = 0 (i.e., for an homogeneous system) the BH Hamiltonian 
is a reasonable description of a system of neutral atoms loaded in optical lattices \cite{gre2,pilati}. However, for low enough $V_{x,y,z}$'s 
\cite{feli4} or thin enough tubes \cite{sine,feli3}, several set of calculations indicate that the results obtained 
from Eq. (\ref{hub}) are different than the ones derived from the full Hamiltonian of Eq. (\ref{hamiltonian}).   

In particular, the superfluid-Mott insulator transition can appear at different values of $V_0$ for continuous and discrete Hamiltonians. A Mott
phase is an incompressible state defined by the condition $\kappa= \partial n/\partial \mu$ = 0, 
where $n$ is the number of particles per potential well, an $\mu$ and $\kappa$ stand for the chemical potential  
and the compressibility of the system as a whole, respectively. For an homogeneous system, 
this condition is only fulfilled when $n$ is
an integer and physically means that adding a single particle to the optical lattice produces a jump in the value of $\mu$.   
In fact, a standard method to know if we have a Mott insulator involves tracking $\mu$ around $n$=1 (or any other integer) and see if there is any 
discontinuity in $\mu$  \cite{feli1,feli2,feli3,feli4,batrou1,batrou2} for increasing values of $V_0$. The critical $V_0$ for the
superfluid-Mott insulator transition is the one below which that discontinuity is absent and above which a jump is clearly seen.
However, this method is only valid for homogeneous systems, in which $\mu$ is the same for every well. 
When the translational invariance is broken, i.e., when $\epsilon_i$ is different for each site $i$,  we can define a local chemical potential, 
$\mu_i$, and from it a local compressibility by \cite{dm,dm2,dm3,dm4}
\begin{equation}\label{kappai}
\kappa_i = \frac{\partial n_i}{\partial \mu_i} 
\end{equation} 
that varies depending on the particular position we are in. This means that $\kappa_i$ can be zero at a particular point, while the
(global) compressibility $\kappa$ is not. The standard definition for $\mu_i$ is \cite{nandini,dm2,dm3,dm4}: 
\begin{equation} \label{mui}
\mu_i = \mu - V_c r_i^2  = \mu - \frac{1}{2} m \omega_z^2 z_i^2,
\end{equation}  
what transforms Eq. (\ref{kappai}) into \cite{nandini,carrasquilla}: 
\begin{equation}\label{kappa2}
\kappa_i =  -\frac{1} {m \omega_z^2 z} \frac{\partial n}{\partial z}.
\end{equation} 
Thus, for analogy to the case of a Mott phase, in which $\kappa$ =0, we can define a Mott domain as the set of contiguous sites for 
which $\kappa_i$ = 0.  By Eq. (\ref{kappa2}), this translates into a set of potential wells with the same number of particles on them. 
This is equivalent to say that in a Mott domain
\begin{equation} \label{delta} 
\Delta_i = < n_i^2> - <n_i>^2 = 0,
\end{equation}
i.e., the local density fluctuations, $\Delta_i$, computed as the variance of the well populations for a set of 100 independent Monte Carlo calculations, are equal to zero.
As we will see, both $\kappa_i$ and $\Delta_i$ behave in a similar way and can be used indistinctly.   

In this work, we study the appearance of Mott domains in systems described by the continuous Hamiltonian of Eq. (\ref{hamiltonian}). 
Three different values of $\omega_z$ and particle numbers, $N$, in the range $N$ = 5-49  were considered. 
We can think of these arrangements as inhomogeneous systems or simply as quasi-one dimensional clusters.  
For each ($\omega_z,N$) combination, 
we obtained a critical value of $V_0$ for the appearance of a Mott domain. For us, this means a set of contiguous potential wells for 
which $\kappa_i$ = 0 (or $\Delta_i$ = 0), and $n_i$ is constant {\em at the same time}. For the number of particles considered in this 
work, this means $n_i$ = 1.

\section{Method}

\begin{figure}
\begin{center}
\includegraphics[width=0.44\textwidth]{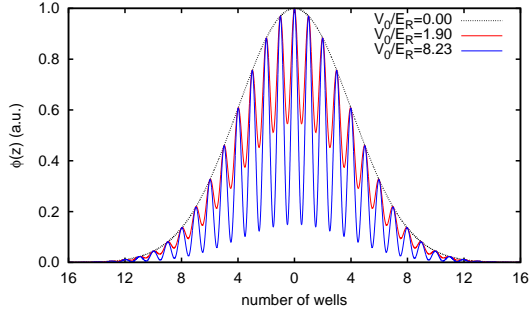} 
\caption{(Color online). $\phi (z)$ for $\omega_z=2 \times \pi 415 Hz$.
}
\label{fig1}
\end{center}
\end{figure}

To solve the Sch\"odinger equation for the Hamiltonian we are interested in, we used the diffusion Monte Carlo (DMC) technique \cite{boro94}. 
This numerical method produces an accurate approximation to the ground state of the system if the initial approximation needed, the so-called   
{\em trial} function, is close enough to the real wavefunction. Since the temperatures at which the experiments are done are very low, the 
ground state is expected to be a reasonable description of the real systems.  The {\em trial} function used in this work is: 
\begin{equation} \label{trialtot}
\Phi({\bf r_1},\cdots,{\bf r_N}) = \prod_{i=1}^N \psi (x_i,y_i) \prod_{j=1}^N \phi (z_i) \prod_{l<m=1}^N \Psi(r_{lm})
\end{equation}
where {\bf r$_i$} are the positions of each of the $N$ neutral atoms in the optical lattice, and $x_i,y_i,z_i$ their
respective coordinates. Here, $\psi (x_i,y_i)$ is the exact solution of the harmonic potential that traps transversally the particles in the 
tube, i.e., a Gaussian of variance $\sigma_{\perp}^2 = \hbar/(m \omega_{\perp})$ (see Eq. (\ref{pot1d})). On the other hand,
\begin{equation} \label{phi}
\phi (z_i)=\exp(-Cz_i^2)\left[1-\alpha \sin^2 \left(\frac{2\pi}{\lambda}z_i\right)\right]
\end{equation}
where $C$ = $(m \omega_z/2 \hbar)$. This makes the first part of $\phi$ simply the exact solution of the longitudinal harmonic oscillator 
when $V_0$ = 0. $\alpha$ is a constant variationally obtained for each combination ($\omega_z,V_0$). 
Examples of $\phi_z (z)$
are displayed in Fig.~(\ref{fig1}), where one can see the maxima around the 
positions of the optical lattice minima.

The remaining part of Eq.(\ref{trialtot}) takes into account the two-body
correlations and was chosen to be 
\cite{boro1}:
\begin{equation}\label{trial}                                    
\Psi(r_{ij})=\left\{
\begin{array}{lr}
0 &  r_{ij} < a\\
B \frac{\sin(\sqrt{\epsilon} (r_{ij}-a))}{r_{ij}} & a<r_{ij}<D,  \\ 
1-A e^{-r_{ij}/\gamma} & r_{ij}>D 
\end{array} 
\right. 
\end{equation}
In principle, this expression depends on five constants, ($A$, $B$, $\epsilon$, $\gamma$ and $D$), that are reduced to two after imposing 
$\Psi(r_{ij})$ and its first and second derivatives to be continuous at $r_{ij}= D$. The undefined the parameters were 
obtained variationally.

To actually perform the calculations, we have to specify all the parameters in Eq. (\ref{hamiltonian}).  
All the energies will be given in units of $E_R$, the so-called recoil energy (($E_R$ = $h^2/2m \lambda^2$)), and the lengths in 
units of $a$. This allows us to get rid of the dependence of the mass in the continuous Hamiltonian. 
The laser wavelength length was fixed to $\lambda_z$ = $\lambda$ = 50$a$, a value
used in previous simulations \cite{troyer,feli1,feli2,feli3,feli4}, while 
the width of the tube $\sigma_{\perp}$ = $(\hbar/m \omega_{\perp})^{1/2}$ was 
set to $\sigma_\perp$  = 3.16$a$. 
As indicated above, three trapping frequencies were 
considered: 60 Hz, 4.15 Hz and 415 Hz. The first was 
taken from the experimental paper by Paredes {\em et al.} \cite{paredes}, and was somehow typical of  these
quasi-one dimensional systems ($\omega_z$ is usually in the range 2$\times\pi$ 20-150 Hz \cite{stoferle,haller,imai}). 
The 4.15 Hz case was 
intended to be something of a lower limit for the longitudinal confinement, chosen to be smaller than the smallest we found in the literature
(9.5 Hz for a two-dimensional optical lattice \cite{gem}). A hundredfold increase in $\omega_z$ was deemed to be sufficient 
as an upper limit for this parameter.         

\section{Results}

As indicated above, we defined a Mott domain as a set of contiguous sites, $i$, for which $\kappa_i$ = 0 (or $\Delta_i$ = 0) and 
$n_i$= 1. 
The number of particles per potential well, $n_i$, has been obtained by integrating the density profiles, $\rho(z)$, in our continuous model, i.e.,  
\begin{equation} \label{histogram}
n_i = \int_{z_i - \lambda/4}^{z_i + \lambda/4}  \rho(z) dz ,
\end{equation}      
where $z_i$ is the position of the center of the potential well $i$ we are interested in, and $\lambda$/4 is the distance from that center to the nearest
maxima of the external potential. The density profiles are obtained averaging up to one hundred independent simulations.

\begin{figure}
\begin{center}
\includegraphics[width=0.44\textwidth]{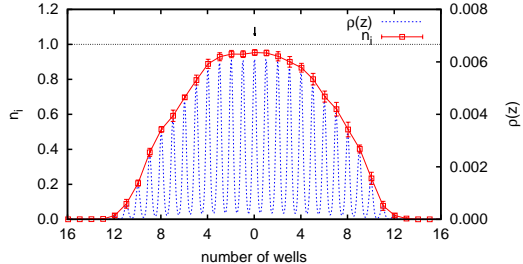} 
\caption{(Color online) Particle density, $\rho(z)$ (dashed line) and number of particles per potential well, $n_i$ (full line) for a set
of $N$= 15 particles, $\omega_z$ = 2$\times\pi$ 415 Hz and $V_0$ = 7.6$E_R$. No Mott domain is observed. An arrow indicates the central well, 
in whose center the longitudinal harmonic potential equals 0.   
}
\label{fig2}
\end{center}
\end{figure}

\begin{figure}
\begin{center}
\includegraphics[width=0.44\textwidth]{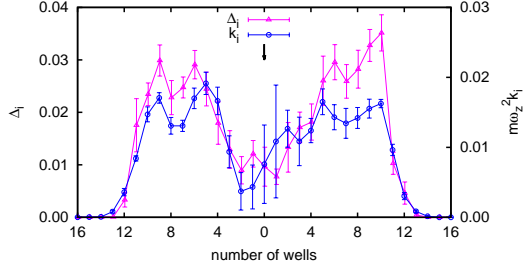} 
\caption{(Color online). $m \omega_z^2 \kappa_i$ (dashed line and triangles) and $\Delta_i$ (full line and circles) for the same system as
in  Fig.~(\ref{fig2}).  
}
\label{fig3}
\end{center}
\end{figure}

Fig.~(\ref{fig2}) displays the density profile and the number of particles per well for a case with small $N$ (15 particles) and low $V_0$ (7.6$E_R$). 
In this profile no Mott domain is present.
That conclusion is supported by the analysis of Fig.~(\ref{fig3}). There, we represent 
$m \omega_z^2 \kappa_i$ (derived form Eq. (\ref{kappa2})) and $\Delta_i$, obtained from Eq. (\ref{delta}) for the same arrangement 
as in Fig.~(\ref{fig2}). $\kappa_i$ was multiplied by $m \omega_z^2$ in order to make 
both magnitudes comparable in the same scale. We can see that none of the conditions to have a Mott domain ($\kappa_i$ = 0 and/or $\Delta_i$ = 0 and $n_i$ =  1 for 
some $i$) are fulfilled. From now on, we will say that cluster such those, with no Mott domains are in State I (phases are not possible in 
inhomogeneous systems). Those arrangements are supposed to be superfluids \cite{dm4}. In Fig.~(\ref{fig2}) we can see also that $\kappa_i$ and $\Delta_i$ behave            
in a similar way. In particular, both of them are different of zero for all $i$'s, and display maxima and minima approximately at the same points. 

\begin{figure}
\begin{center}
\includegraphics[width=0.44\textwidth]{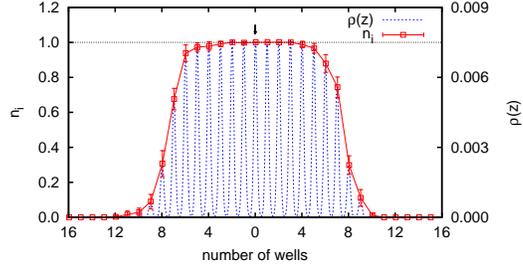} 
\caption{(Color online) Same as in Fig.~(\ref{fig2}) but for $V_0$ = 15.2$E_R$. 
}
\label{fig4}
\end{center}
\end{figure}

\begin{figure}
\begin{center}
\includegraphics[width=0.44\textwidth]{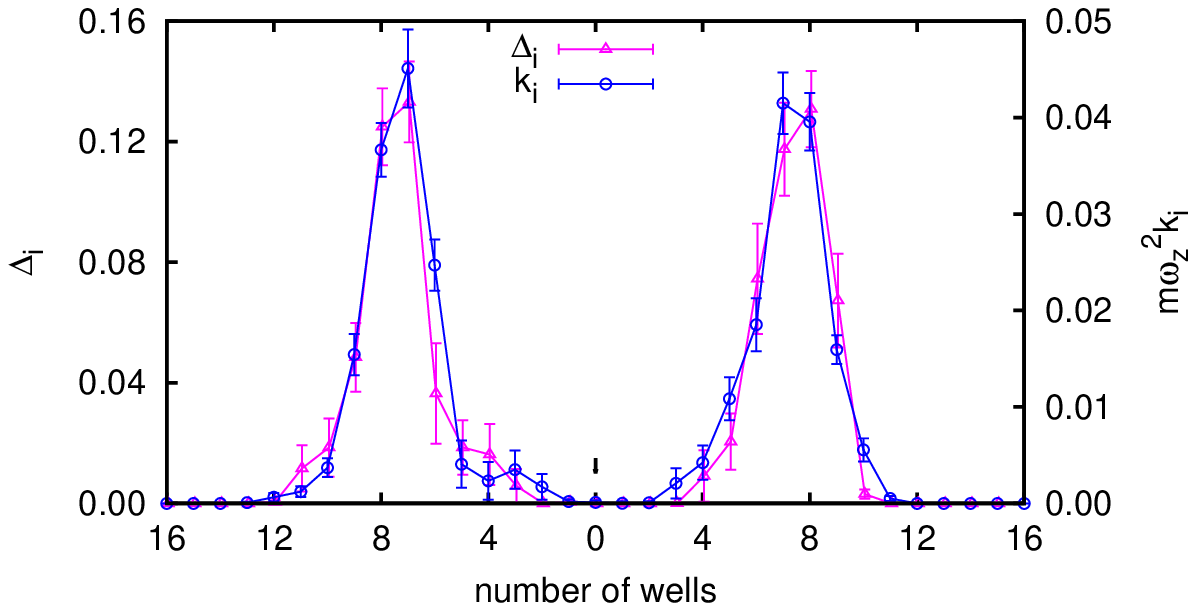} 
\caption{(Color online) Same as in Fig.~(\ref{fig3}) but for $V_0$ = 15.2$E_R$.
}
\label{fig5}
\end{center}
\end{figure}

The situation changes when we increase $V_0$ while keeping constant the rest of the parameters. Then, the average 
number of atoms at the central well(s) increases steadily up to $n_i$ = 1. An example of this new situation is displayed 
in Fig.~(\ref{fig4}) for $N$ = 15, $f_z$ =  415 Hz and  $V_0$ = 15.2$E_R$. There, we can see a plateau
around $i$=0 (indicated by an arrow), in which $n_i$ = 1, i.e.,  
a Mott domain. With the help of Fig.~(\ref{fig5}) we can see also that, in the same set of sites, $\kappa_i$ = $\Delta_i$ = 0.
Clusters with only one insulating domain are considered in the following to be in State II.
In going from a cluster in State I to a cluster in State II, there is  
a value of $V_0$ above which, within two standard deviations of the reference values,  
$n_i$ = 1 and $\kappa_i$ = $\Delta_i$ = 0 at the same time, for at least one of the three central wells of the optical lattice.   
We call that critical value $(V_0)_C$.
For the case depicted in Figs.(\ref{fig2})-(\ref{fig5}) the critical value for the transition between State I and State II was  $(V_0/E_R)_C$ = 8.2 $\pm$ 0.6. 
With a similar procedure, we can obtain a set of triads  
($\omega_z,N,(V_0)_C$) that define the state diagram of the system \cite{dm4}. No phase diagram can be obtained since, as the system is
inhomogeneous, even for very large values 
of $V_0$, there are always non insulating "wings" in the regions further from the center for which $n_i \neq$ 1 \cite{dm,dm2,dm3,dm4}.                

\begin{figure}
\begin{center}
\includegraphics[width=0.44\textwidth]{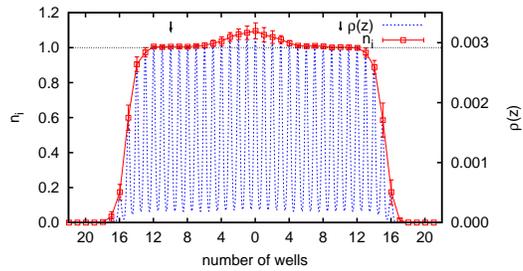} 
\caption{(Color online) Same as in Fig.~(\ref{fig2}) but for $N$ = 31 and $V_0$ = 6.3$E_R$.
}
\label{fig6}
\end{center}
\end{figure}

For $N$ and/or $f_z$ small enough, the only possible profiles are similar either to that of Fig.~(\ref{fig2}) or of Fig.~(\ref{fig4}). i.e.,
either we have a Mott domain at the center of the trap or we have not. On the other hand, when $N$ and/or $f_z$ are large enough, we have situations as the 
one depicted in Fig.~(\ref{fig6}). Those clusters are said to be in State III, and  
they have two Mott domains symmetrically located around $i=$ 0.  
The centers of those 
domains are signaled by two downward pointing arrows. 
When $V_0$ increases further, the system ends ups in a situation similar to the depicted in
Fig.~(\ref{fig4}): a single-connected Mott domain that covers most of the system. We have then two critical values of $V_0$: one for the appearance 
of the two separated Mott plateaus, and another (and larger), for the creation of a single Mott domain. This kind of shell structure has been 
experimentally observed \cite{shell}.  

\begin{figure}
\begin{center}
\includegraphics[width=0.44\textwidth]{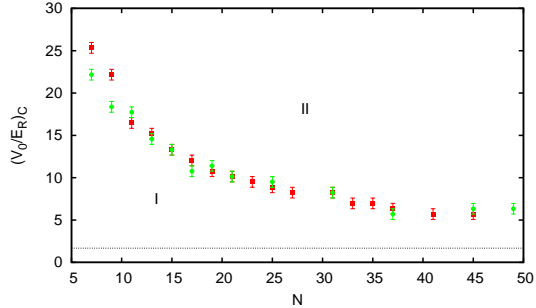} 
\caption{(Color online) Critical values of $(V_0/E_R)_C$ for the apparition of a Mott domain in the center of 
cluster, in terms of the number of particles ($N$) for $N$ = 5-49.
Squares,  $\omega_z$ = 2$\times \pi$ 4.15 Hz; circles, $\omega_z$ = 2$\times \pi$ 60 Hz. The dotted line is the value 
for the homogeneous case with the same optical lattice parameters, taken from Ref. \onlinecite{feli3}.
}
\label{main1}
\end{center}
\end{figure}

In Fig.~(\ref{main1}) we display the state diagram for  $\omega_z$ = 2$\times \pi$ 4.15 Hz (squares) and  $\omega_z$ = 2$\times \pi$ 60 Hz (circles),
for numbers of particles in the range $N$ = 5-49. Under those conditions, we have only clusters in State I or State II. 
We can see that in both curves 
the critical value of the potential well necessary to create a cluster in State II decreases with $N$ with little difference between both sets of data. 
Also displayed is the critical $V_0$ value for an homogeneous system with the same $\sigma_{\perp}$ and $\lambda$
($(V_0/E_R)_C$ = 1.7 $\pm$ 0.3, Ref. \onlinecite{feli3}), noticeably lower than the values for any of the cluster values represented in Fig.~(\ref{main1}).
We can see also that $(V_0/E_R)_C$ seems to level off for clusters with $N >$ 40, to a number more than twice as the corresponding to the equivalent 
homogeneous system.       

\begin{figure}
\begin{center}
\includegraphics[width=0.44\textwidth]{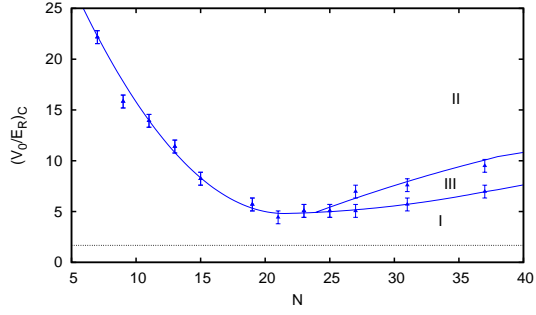} 
\caption{(Color online)  
Same as in Fig. (\ref{main1}), but for $\omega_z$ = 2$\times \pi$ 415 Hz and $N$= 5-37.  
}
\label{main2}
\end{center}
\end{figure}
 
Fig.~(\ref{main2}) gives us the same information as Fig.~(\ref{main1}) but for 
$\omega_z$ = 2$\times \pi$ 415 Hz. We can see that the curve is similar to  
one in the previous figure up to $N$= 20.   
A further increase in the number of atoms loaded in the optical lattice makes the critical value for the disappearance of State I
grow again. However, when this happens, the transition is not to State II as in the previous cases, but to State III. A further increase in $V_0$ is necessary 
to produce a cluster in State II. This second set of critical values, higher than the previous ones, 
for the change State III $\rightarrow$ State II is also displayed.                           

\begin{figure}
\begin{center}
\includegraphics[width=0.44\textwidth]{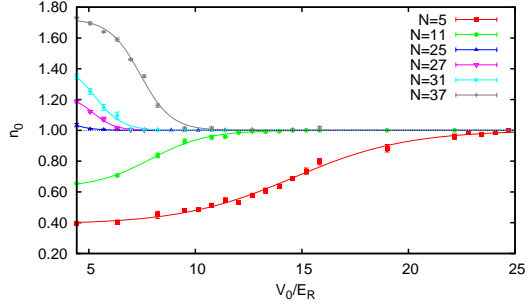} 
\caption{(Color online)  
Number of particles at the central site, $i$ =0, as a function of $V_0/E_R$ for different values of $N$ and $\omega_z$ = 2$\times \pi$ 415 Hz.  
}
\label{fig10}
\end{center}
\end{figure}

The appearance of State III for certain values of $N$ and $V_0$ can be understood with the help of Fig.~(\ref{fig10}). There, we plot the evolution of the number of particles on the central well, $n_0$,
as a function of the  external potential depth for $f_z$ = 415 Hz. We can see that for small values of both $N$ and $V_0$, that occupation is smaller than one, and grows with $V_0$ to reach $n_0=$1, 
as corresponds to a Mott insulator domain.  
On the other hand, when $N$ is larger, $n_0 >$1 for $V_0 \rightarrow 0$ and reaches unity, as before, for large $V_0$ values. The limit between those regimes corresponds to $N \sim$ 25; for larger $N$ values we can have State III clusters.  
The fact that for $\omega_z$ = 2 $\pi \times$ 4.15 and $\omega_z$ = 2 $\pi \times$ 60 Hz, 
$n_0$ is always less than one in the limit $V_0 \rightarrow$ 0 for all the values of $N$  considered in this work, suggests that a necessary condition to see State III clusters is that the number of
particles on the central well be at least one for low enough values of the potential depth.  

\section{Discussion}

If we look at Figs.~(\ref{fig2})-(\ref{fig6}), 
we find that the profiles displayed there
are similar to the ones found in the literature for one-dimensional Bose-Hubbard Hamiltonians \cite{dm,dm2,dm3,dm4}. Then, it would 
appear that continuous model calculations could only certify the validity of that discrete approximation, at much higher computational cost.
In any case, our data have at least an advantage with respect to the ones derived form a BH model: the state diagrams depend directly on the experimental parameters  
($V_0,\lambda,\omega_z,\omega_{\perp}$) and the results do not need any translation from the $J,U$ and $V_c$ parameters to the real ones 
via Eq. (\ref{J}),(\ref{U}) and (\ref{epsilon}).  

However, further analysis indicates that our results are not equivalent to the ones obtained from a BH Hamiltonian. In particular, in Refs. \onlinecite{dm4} and \onlinecite{dm5} 
is shown that, due to scaling arguments, the density profiles depend only on a reduced variable, $\tilde\rho = N \sqrt{(V_c/J)}$. Since those profiles are used 
to derive the state diagrams, the critical $V_0$'s should depend only on that variable. Contrarily to what happens in the BH description, this is not true in our simulation. 
For instance, in Fig.~(\ref{main1}) we can see that the results for two different trappings are virtually on top of each other, instead of depending on the corresponding $\omega_z$'s.     
Moreover, all trials to reduce the three curves presented in Figs.~(\ref{main1}) and Fig.~(\ref{main2}) to a single one have been unsuccessful.


In a pure one-dimensional BH Hamiltonian, to have a Mott domain 
we need $U/J \ge$ 5.5 \cite{dm4}, a larger value than the corresponding to a homogeneous, non-trapped system ($U/J \sim$  3.6).  
$U/J$ = 5.5 translates into $V_0/E_R$ = 1.6 (Eqs. (\ref{J}) and (\ref{U})), and larger values of $U/J$ would also turn into  $V_0/E_R$'s greater than 1.6. All this means that a minimum $U/J$ implies 
the existence of a minimum $V_0/E_R$  below which we can have only superfluid clusters. 
This feature can be seen clearly in Fig. \ref{main2}, in which the minimum is  $V_0/E_R \sim$ 4.4. The plateau observed in Fig. \ref{main1} suggests 
that this is also the case for smaller confinements, with superfluid clusters for $V_0/E_R < $ 5.7. Both values are larger than the $(V_0/E_R)_C$ deduced from the BH state diagram ($V_0/E_R$ = 1.6,
given above). i.e., a BH model underestimates the $V_0$ value needed to have a Mott domain with respects to the results from a continuous Hamiltonian. 
In this, a trapped system is similar to an homogeneous one, in which 
$(V_0/E_R)_{BH}$ = 0.7 $< (V_0/E_R)_{HS}$ =1.7 $\pm$ 0.3 \cite{feli3}.  

\acknowledgments

We acknowledge partial financial support from the 
Junta de Andaluc\'{\i}a group PAI-205 and grant FQM-5987, DGI (Spain) grant No. FIS2010-18356.


\end{document}